\documentclass[runningheads]{llncs}
\usepackage{graphicx}
\usepackage{bibnames}
\usepackage{amsmath}
\usepackage{mathrsfs}
\usepackage{amsfonts}
\usepackage{mathtools}

\usepackage{xcolor}
\usepackage[ruled]{algorithm2e}

\SetAlFnt{\small}
\SetAlCapFnt{\small}
\SetAlCapNameFnt{\small}
\SetAlCapHSkip{0pt}
\IncMargin{-\parindent}

\begin{document}
\title{Bayesian Persuasion in Sequential Trials}
%
%
\author{Shih-Tang Su\inst{1}\thanks{Supported in part by NSF grant ECCS 2038416 and MCubed 3.0.} \and
Vijay G. Subramanian\inst{1}\thanks{Supported in part by NSF grants ECCS 2038416, CCF 2008130, and CNS 1955777.} \and
Grant Schoenebeck\inst{1}\thanks{Supported in part by NSF grants CCF 2007256 and CAREER 1452915.}}
%
\authorrunning{S. Su et al.}
%
\institute{University of Michigan, Ann Arbor \\ \email{shihtang@umich.edu}, \email{vgsubram@umich.edu}, \email{schoeneb@umich.edu}}
\maketitle              
\begin{abstract}
We consider a Bayesian persuasion problem where the sender tries to persuade the receiver to take a particular action via a sequence of signals. This we model by considering multi-phase trials with different experiments conducted based on the outcomes of prior experiments. 
In contrast to most of the literature, we consider the problem with constraints on signals imposed on the sender. This we achieve by fixing some of the experiments in an exogenous manner; these are called determined experiments. 
This modeling helps us understand real-world situations where this occurs: e.g., multi-phase drug trials where the FDA determines some of the experiments, start-up acquisition by big firms where late-stage assessments are determined by the potential acquirer, multi-round job interviews where the candidates signal initially by presenting their qualifications but the rest of the screening procedures are determined by the interviewer. 
The non-determined experiments (signals) in the multi-phase trial are to be chosen by the sender in order to persuade the receiver best. With a binary state of the world, we start by deriving the optimal signaling policy in the only non-trivial configuration of a two-phase trial with binary-outcome experiments. We then generalize to multi-phase trials with binary-outcome experiments where the determined experiments can be placed at arbitrary nodes in the trial tree. Here we present a dynamic programming algorithm to derive the optimal signaling policy that uses the two-phase trial solution's structural insights. We also contrast the optimal signaling policy structure with classical Bayesian persuasion strategies to highlight the impact of the signaling constraints on the sender.

\end{abstract}

\vspace{-16pt}
\section{Introduction} \label{intro}
\vspace{-4pt}
Information design studies how informed agents (senders) persuade uninformed agents (receivers) to take specific actions by influencing the uninformed agents' beliefs via information disclosure in a game. The canonical Kamenica-Gentzkow model \cite{kamenica2011bayesian} is one where the sender can commit to an information disclosure policy (signaling strategy) before learning the true state. Once the state is realized, a corresponding (randomized) signal is sent to the receiver. Then, the receiver takes an action, which results in payoffs for both the sender and the receiver. Senders in information design problems only need to manipulate the receivers' beliefs with properly chosen signals. The manipulated beliefs will create the right incentives for the receiver to spontaneously take specific actions that benefit the sender (in expectation). In (classical) mechanism design, however, the story is different: the designer is unaware of the agents' private information, and the agents communicate their private information to the designer, who then has to provide incentives via (monetary) transfers or other means. The flexibility afforded by information design that allows the sender to benefit from information disclosure without implementing utility-transfer mechanisms has led to greater applicability of the methodology: various models and theories can be found in survey papers such as \cite{bergemann2019information} and \cite{kamenica2019bayesian}.

Our work is motivated by many real-world problems where persuasion schemes are applicable, but the sender is constrained in the choice of signals available for information design. Specifically, we are interested in problems that are naturally modeled via multi-phase trials where the interim outcomes determine the subsequent experiments. Further, we insist that some of the experiments are given in an exogenous manner. This feature imposes restrictions on the sender's signaling space, and without it, we would have a classical Bayesian persuasion problem with an enlarged signal space. Our goal is to study the impact of such constraints on the optimal signaling scheme, and in particular, to contrast it with the optimal signaling schemes in classical Bayesian persuasion.

The following motivating example describes a possible real-world scenario.
\vspace{-4pt}
\begin{example}[Motivating example - Acquiring funds from a venture capital firm]
We consider a scenario where a start-up is seeking funds from a venture capital firm. The process for this will typically involve multiple rounds of negotiation and evaluation: some of these will be demonstrations of the start-up's core business idea, and the others will be assessments by the venture capital firm following their own screening procedures. The start-up will have to follow the venture capital firm's screening procedures but chooses its product demonstrations. Based on these stipulations, the start-up needs to design its demonstrations to maximize its chance of getting funded.
\end{example}
\begin{figure}[ht] \vspace{-18pt}
    \centering
    \includegraphics[width=0.9\textwidth]{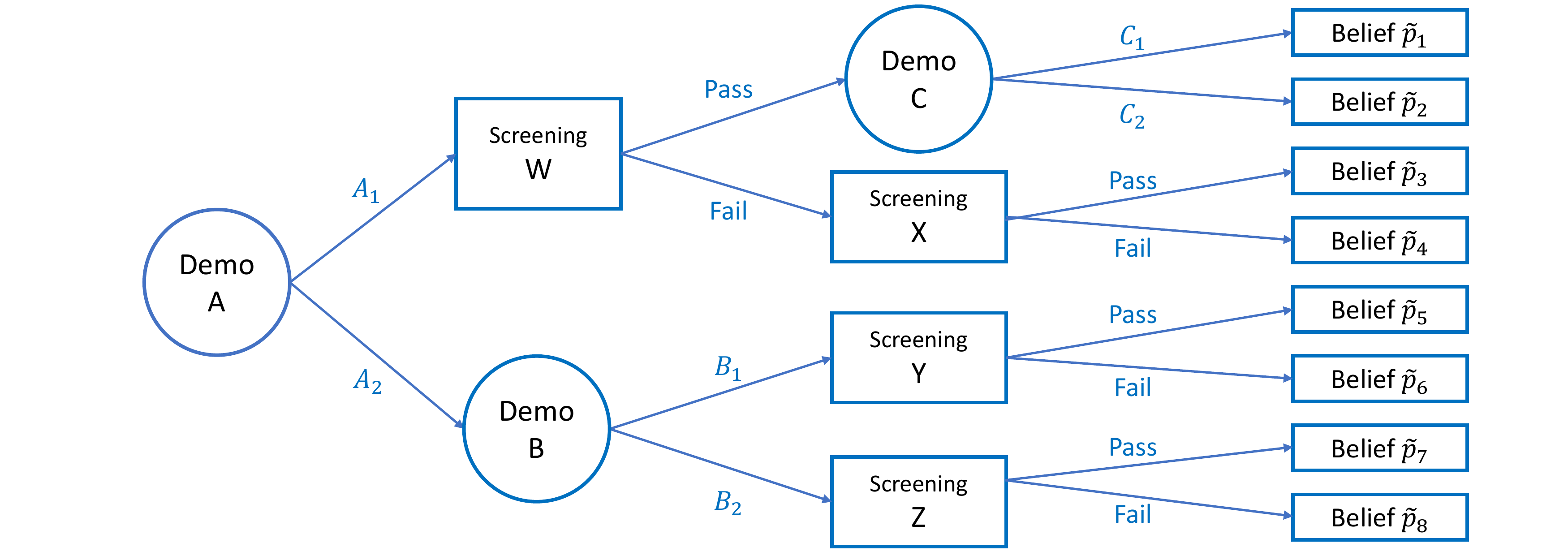}
     \vspace{-6pt}
    \caption{Example of a negotiation process -- a startup vs. a venture capital firm.}
    \label{fig:motivate} \vspace{-12pt}
\end{figure}
In the example above, the start-up (sender) has to generate an information disclosure scheme to get the desired funds from the venture capital firm (receiver). Then the screening procedures set by the venture capital firm are analogous to our determined experiments, and the demonstrations carried out by the start-up are the (sender) designed experiments. For example, in Figure~\ref{fig:motivate} we present one plausible interaction where the start-up company designs demonstrations A, B, and C (circles in the figure) and the venture capital firm has pre-determined screening examinations W, X, Y, and Z (rectangles in the figure). 
Whereas we have illustrated this example via a balanced tree, if we have an unbalanced tree owing to the receiver deciding in the middle, we can modify it to a balanced tree by adding the required number of dummy stages.

The sender's reduced flexibility on her signaling strategies under some predetermined experiments with arbitrary positions and informativeness differentiates our work from the growing literature on dynamic information design. Our model considers a problem with the following features: a static state space, a sequential information disclosure environment, and a signaling space restricted by some exogenous constraints whose harshness may depend on the proposed singling schemes. Models with a static state space, an unrestricted signal space but a variety of sequential information disclosure environments have been studied to capture features in different real-world problems: e.g., with multiple senders \cite{forges2008long,li2018bayesian}, with costly communication \cite{honryo2018dynamic,nguyen2019bayesian}, allowing for sequential decision making \cite{ely2020moving}, or with partial commitment \cite{au2015dynamic,nguyen2019bayesian}. Models with dynamic states and sequential information disclosure environments are usually studied under an informed sender with the knowledge of dynamically changing state(s); a variety of works in this category lie in state change detection \cite{ely2017beeps,farhadi2020dynamic} or routing games \cite{meigs2020optimal}. Although several works \cite{dughmi2016persuasion,gradwohl2021algorithms,le2019persuasion} also consider constrained signaling schemes, these works either consider the signal space to be smaller than the action space \cite{gradwohl2021algorithms,dughmi2016persuasion} or consider a noisy signaling environment \cite{le2019persuasion}. Models with exogenous information \cite{kolotilin:mylovanov:zapechelnyuk:li:persuasion:2017,bizzotto2021can,bizzotto2021dynamic}, can be viewed as sequential information disclosure problems with exogenous determined experiments placed in determined phases. The sequential information disclosure in our model, which actually enlarges the signal space, makes our work different from above works. 
To keep the focus of the paper on sequential trials, we discuss the broader literature on constrained senders, algorithmic information design, and works related to the receiver's experiment design\footnote{See Section~5 in \cite{our_arxiv} for details.} in our online version \cite{our_arxiv}. 



According to the motivating example illustrated in Figure~\ref{fig:motivate}, the persuasion problem considers a sequence of experiments where experiments further along in the tree depend on the outcomes of previous phases. The experiment to be run in each phase is either exogenously determined or chosen by the sender. In the game, the sender chooses designed experiments with knowledge of the prior, the determined experiments, and the receiver's utility function, but before the state of the world is realized. After the sender commits to the experiments (i.e., the signaling strategy), the state of the world is realized, and a specific sequence of experiments is conducted based on the realization of the underlying random variables. The receiver then takes an action depending on the entire sequence of outcomes. The prior, the sender's and receiver's utility functions, the determined experiments, and the designed experiments (after the sender finalizes them) are assumed to be common knowledge. We study this problem for binary states of the world, first for two-phase binary-outcome trials, and then generalized to multi-phase binary-outcome trials. We then generalize to non-binary experiments (still with an underlying binary state space).  In the online version~\cite{our_arxiv} we add games with an additional stage where the receiver moves before the sender to decide some or all of the determined experiments, perhaps with some constraints.

\vspace{4pt}
\noindent\textbf{Contributions:}  The main contributions of this work are:
\vspace{-4pt}

\begin{enumerate}
    \item To the best of our knowledge, within the multi-phase Bayesian persuasion framework, we are the first to study the design of sender's signaling schemes with exogeneously determined experiments in arbitrary positions. Our results highlight the difference between ``exogenously determined experiments'' and ``exogenously given information'' in the dynamic information design, where the former gives greater flexibility and allows for more heterogeneity.
    \item We explicitly solve the optimal signaling scheme in two-phase trials. Moreover,
    using structural insights gained from two-phase trials, we present a dynamic programming algorithm to derive the optimal signaling in general multi-phase trials via backward iteration.
    \item We analyze the impact of constraints on the sender via the determined experiments by contrasting the performance with the classical Bayesian persuasion setting and when using classical Bayesian persuasion optimal signaling schemes when the sender is constrained. As a part of this, we provide a sufficient condition for when a sequential trial is equivalent to classical Bayesian persuasion with a potentially enlarged signal space.
\end{enumerate}

\vspace{-18pt}
\section{Problem Formulation} \label{problem}
\vspace{-6pt}
There are two agents, a sender (Alice) and a receiver (Bob), participating in the game. We assume binary states of the world, $\Theta=\{\theta_1,\theta_2\}$, with a prior belief $p \coloneqq \mathbb{P}(\theta_1)$ known to both agents. The receiver has to take an action $\Phi \in \{\phi_1,\phi_2\}$ which can be thought of as a prediction of the true state.  We assume that the receiver's utility
is given by $u_r(\phi_i,\theta_j)=1_{\{i=j\}}$ for all $i,j\in \{1,2\}$. To preclude discussions on trivial cases
and to simplify the analysis, we assume that the sender always prefers the action $\phi_1$, and her utility is assumed to be $u_s(\phi_1,\theta_i)=1$ and $u_s(
\phi_2,\theta_i)=0$ for all $i\in \{1,2\}$. 

Before the receiver takes his action, a trial consisting of multiple phases will be run, and the outcome in each phase will be revealed to him. In each phase, one experiment will be conducted, which is chosen according to the outcomes in earlier phases. Hence, the experiment outcomes in earlier phases not only affect the interim belief but also influence the possible (sequence of) experiments that will be conducted afterward. 
In the most sender-friendly setup where the sender can choose any experiment in each phase without any constraints, the problem is equivalent to the classical Bayesian persuasion problem with an enlarged signal space. However, when some experiments are pre-determined conditional on a set of outcomes, the sender must take these constraints into account to design her optimal signaling structure. 

To present our results on the influence of multiple phases on the sender's signaling strategy, we start with a model of two-phase trials with binary-outcome experiments in the rest of this section. We then analyze the optimal signaling strategy of this model in Section~\ref{sec:BOE}. After that, we will introduce the general model of multiple-phase trials with binary-outcome experiments and propose a systematic approach to analyze the optimal signaling structure in Section~\ref{sec:multiphase}. 

\vspace{-8pt}
\subsection{Model of Binary-Outcome Experiments in Two-Phase Trials} \label{sec:basicmodel}
\vspace{-4pt}
There are two phases in the trial: phase I and phase II. Unlike in the classical Bayesian persuasion problem, our goal is for the sender to not have the ability to choose the experiments to be conducted in both phases of the trial. We will start by assuming that the sender can choose any binary-outcome experiment in phase I, but both the phase-II experiments (corresponding to the possible outcomes in phase I) are determined. Formally, in phase I, there is a binary-outcome experiment with two possible outcomes $\omega_1\in \Omega_1=\{\omega_{A},\omega_{B}\}$ and each outcome corresponds to a determined binary-outcome experiment, $E_A$ or $E_B$, which will be conducted in phase II, respectively. The sender can design the experiment in phase I via choosing a probability pair $(p_1,p_2)\in[0,1]^2$, where $p_i=\mathbb{P}(\omega_{A}|\theta_i)$. Once the probability pair $(p_1,p_2)$ is chosen, the interim belief of the true state $\mathbb{P}(\theta_1|\omega_1)$ can be calculated (while respecting the prior) as follows:
\begin{align}
\label{eq:interimbelief}
\mathbb{P}(\theta_1|\omega_A)=\frac{pp_1}{pp_1+(1-p)p_2} \text{ and }
\mathbb{P}(\theta_1|\omega_B)=\frac{p(1-p_1)}{p(1-p_1)+(1-p)(1-p_2)}.
\end{align}

On the other hand, the phase-II experiments are given in an exogenous manner beyond the sender's control. In phase II, one of the binary-outcome experiments, $E \in \{E_A, E_B\}$ will be conducted according to the outcome,  $\omega_A$ or $\omega_B$ of the phase-I experiment.  If  $\omega_A$ is realized, then experiment $E_A$ will be conducted in phase II; if  $\omega_B$ is realized, the experiment $E_B$ will be conducted in phase II. Similarly, we can denote the possible outcomes $\omega_{2}\in \Omega_{2X}=\{\omega_{XP},\omega_{XF}\}$ when the experiment $E_X$ is conducted, where notation $P, F$ can be interpreted as passing or failing the experiment. Likewise, the phase-II experiments can be represented by two probability pairs $E_1=(q_{A1},q_{A2})$ and $E_2=(q_{B1},q_{B2})$, where $q_{Xi}$ denotes the probability that the outcome $\omega_{XP}$ is realized conditional on the experiment $E_X$ and the state $\theta_i$, i.e., $q_{Xi}=\mathbb{P}(\omega_{XP}|\theta_i,E_X)$.

    In real-world problems, regulations, physical constraints, and natural limits are usually known to both the sender and the receiver before the game starts. Hence, we assume that the possible experiments $E_1,E_2$ that will be conducted in phase II are common knowledge. Given the pairs $(q_{A1},q_{A2}),(q_{B1},q_{B2})$, the sender's objective is to maximize her expected utility by manipulating the posterior belief (of state $\theta_1$) in each possible outcome of phase II. However, since the phase II experiments are predetermined, the sender can only indirectly manipulate the posterior belief by designing the probability pair $(p_1,p_2)$ of the phase-I experiment. As the sender prefers the action $\phi_1$ irrespective of the true state, her objective is to select an optimal probability pair $(p_1,p_2)$ to maximize the total probability that the receiver is willing to take action $\phi_1$. Recalling the receiver's utility function discussed above, the receiver's objective is to maximize the probability of the scenarios where the action index matches the state index. 
Thus, the receiver will take action $\phi_1$ if the posterior belief $\mathbb{P}(\theta_i|\omega)\geq \frac{1}{2}$ and take action $\phi_2$ otherwise.
    After taking the receiver's objective into the account, the sender's optimization problem can be formulated as below:
    \begin{eqnarray} \label{eqn:obj}
        &&\max_{p_1,p_2} \sum_{\omega_2\in\{\omega_{AP},\omega_{AF},\omega_{BP},\omega_{BF}\} }\mathbb{P}(\phi_1,\omega_2)\\
        \text{s.t.} &&\Big(\mathbb{P}\big(\theta_1|\omega_{AY},q_{A1},q_{A2},p_1,p_2\big)-\tfrac{1}{2}\Big)\Big(\mathbb{P}(\phi_1,\omega_{AY})-\tfrac{1}{2}\Big)\geq 0~\forall~ Y\in\{P,F\},\nonumber\\
        &&\Big(\mathbb{P}\big(\theta_1|\omega_{BY},q_{B1},q_{B2},p_1,p_2\big)-\tfrac{1}{2}\Big)\Big(\mathbb{P}(\phi_1,\omega_{BY})-\tfrac{1}{2}\Big)\geq 0~\forall~ Y\in\{P,F\},\nonumber\\
        && \mathbb{P}(\omega_{AP})=pp_1q_{A1}+(1-p)p_2q_{A2}, \nonumber \\ && 
        \mathbb{P}(\omega_{BP})=p(1-p_1)q_{B1}+(1-p)(1-p_2)q_{B2}, \nonumber \\
        && \mathbb{P}(\omega_{AP})+\mathbb{P}(\omega_{AF})=pp_1+(1-p)p_2,  \nonumber \\ && 
        \mathbb{P}(\omega_{BP})+\mathbb{P}(\omega_{BF})=p(1-p_1)+(1-p)(1-p_2),   \nonumber \\
       &&\mathbb{P}(\phi_1,\omega_2)\in [0,1]~\forall \omega_2\in\{\omega_{AP},\omega_{AF},\omega_{BP},\omega_{BF}\}, \qquad 
       p_1,p_2\in [0,1].   \nonumber 
    \end{eqnarray}
    In the sender's optimization problem (\ref{eqn:obj}), the first two inequalities are constraints of incentive-compatibility (IC) that preclude the receiver's deviation. The IC constraints can be satisfied when both terms in the brackets are positive or negative. That is to say; the sender can only persuade the receiver to take action $\phi_1$/$\phi_2$ when the posterior belief (of $\theta_1$) is above/below $0.5$. While we have written the IC constraints in a nonlinear form for compact presentation, in reality they're linear constraints. The next four equations are constraints that make the sender's commitment (signaling strategy) Bayes plausible\footnote{A commitment is Bayes-plausible \cite{kamenica2011bayesian} if the expected posterior probability of each state equals its prior probability, i.e., $\sum_{\omega\in \Omega}\mathbb{P}(\omega)\mathbb{P}(\theta_i|\omega)=\mathbb{P}(\theta_i)$.}. Hence, there are $4$ IC constraints and $4$ Bayes-plausible constraints in the optimization problem for a two-phase trial. However, in an $N$-phase trial, both the number of IC constraints and the number of Bayes-plausible constraints will expand to $2^N$ each. Although the linear programming (LP) approach can solve this optimization problem, solving this LP problem in large Bayesian persuasion problems can be computationally hard \cite{dughmi2019algorithmic}.
    Hence, instead of solving this optimization problem via an LP, we aim to leverage structural insights discovered in the problem to derive the sender's optimal signaling structure.

    We end this section by emphasizing that this model is the only non-trivial two-phase trial configuration when determined and designed experiments coexist. In other configurations such that some of the phase-II experiments can be designed by sender, the model can be reduced to a corresponding single-phase trial in the sense that the single-phase trial will yield the same payoffs for both sender and receiver when they play optimally. (Note that the reduced model may have a different prior if the experiment in phase-I is determined).

\section{Binary-outcome Experiments in Two-phase Trials} \label{sec:BOE}
\vspace{-6pt}
  In this section the sender's optimization problem presented in \eqref{eqn:obj} Section \ref{sec:basicmodel}, 
  is solved starting with the simplest non-trivial case. There are only two phases in the trial studied here, and from this we will develop more insight into how different types of experiments (determined versus sender-designed) influence the optimal signaling strategy of the sender. To be more specific, we will analyze how two determined experiments (in phase II) and one sender-designed experiment (in phase I) will impact the sender's optimal signaling strategy. Before we present the general case, we discuss a subset class of two-phase trials that are similar to single-phase trials. In this class of two-phase trials, in one of the phase-II experiments, called a \textit{trivial} experiment, the outcome distribution is independent of the true state. 
Trivial experiments \cite{basu1975statistical}, also called (Blackwell) non-informative experiments in some literature, are frequently used as benchmarks to compare the agents' expected utility change under different signaling schemes/mechanisms, e.g., \cite{nguyen2019bayesian,li2016blackwell,meigs2020optimal}. 
  This two-phase model with a trivial experiment tries to capture real-world problems with one actual (and costly) experiment, e.g., clinical trials, venture capital investments, or space missions. Since the experiment is costly, a screening procedure is provided to decide whether it is worth conducting the experiment. We will then analyze the optimal signaling strategy in the general scenario, where both experiments in phase II are \textit{non-trivial}.

\vspace{-12pt}    
\subsection{Experiments with screenings} \label{sec:sde}
\vspace{-4pt}   
We start by analyzing the sender's optimal strategy (signaling structure) in a simple scenario where there is one \textit{non-trivial} experiment conducted in phase II. 
The sender's authority on choosing the probability pair $(p_1,p_2)$ controls the screening process. To avoid any ambiguity, we first define what a \textit{trivial} experiment is.
\vspace{-4pt} 
\begin{definition}
    An experiment $E$ is trivial if the distribution of its outcomes $\triangle_E$ is independent of the state of the world: $\triangle_E = \triangle_{E|\theta_i}$ for all $\theta_i\in \Theta$. 
\end{definition}

When a trivial experiment (in phase II) is conducted, the posterior belief of the state stays the same as the interim belief derived in \eqref{eq:interimbelief}. When there exists a trivial experiment in the two phase-II trial options, then Lemma~\ref{lem:screening} states that the sender and the receiver's expected utility under the optimal signaling strategy is the same as in the (single-phase) classical Bayesian persuasion problem.
\vspace{-4pt} 
 \begin{lemma} \label{lem:screening}
    When the state space is binary, both sender and receiver's expected utilities are the same in the following two Bayesian persuasion schemes under each scheme's optimal signaling strategy: \vspace{-6pt}
    \begin{enumerate}
        \item Bayesian persuasion in a single-phase trial,
        \item Bayesian persuasion in a two-phase trial with a sender-designed phase-I experiment and a trivial experiment in phase II.
    \end{enumerate}
\end{lemma}
\vspace{-4pt} 
In the single-trial classical Bayesian persuasion setting, the optimal signaling strategy only mixes the two possible states in one outcome (e.g., when the prosecutor claims the suspect is guilty). On the other outcome, the sender reveals the true state with probability one (e.g., when the prosecutor says the suspect is innocent).
When there is a trivial experiment in phase II, the other experiment (supposing that it will be conducted at outcome $\omega_B$) will be rendered defunct 
by the sender's choice of experiments in phase I. 
This phenomenon occurs because the sender can always choose to reveal the true state when the non-trivial experiment is to be conducted, i.e., by setting $\mathbb{P}(\theta_1|E_B)=1$ or $\mathbb{P}(\theta_2|E_B)=1$; and the classical Bayesian persuasion strategy can be replicated. In essence, having a trivial experiment in the phase-II trial does not constrain the sender.
\vspace{-8pt}
\subsection{Assumptions and induced strategies} \label{sec:twophase}

 Next we detail the optimal signaling strategy in our two-phase trial setting with general binary-outcome experiments. To aid in the presentation and to avoid repetition, we make two assumptions without loss of generality and introduce several explanatory concepts before the analysis.

\begin{lemma} \label{clm_asp}
We can make the following two assumptions WLOG.
 \vspace{-4pt}
\begin{enumerate}
    \item The probability of passing a phase-II experiment under $\theta_1$ is greater than or equal to the probability under $\theta_2$, i.e., $q_{A1}\geq q_{A2}$ and $q_{B1}\geq q_{B2}$.
    \item When the true state is $\theta_1$, the experiment conducted when outcome $\omega_A$ occurs is more informative\footnote{In terms of the Blackwell informativeness from \cite{blackwell1953equivalent}.} than the experiment conducted when outcome $\omega_B$ occurs, i.e., $q_{A1}\geq q_{B1}$.
\end{enumerate}
\end{lemma}

The sender's strategy consists of the following: choice of phase-I experiment parameters $(p_1,p_2)$ and the persuasion strategies in phase-II for each outcome of the phase-I experiment. To understand better the choices available to the sender and her reasoning in determining her best strategy, we will study the possible persuasion strategies in phase-II; these will be called induced strategies to distinguish them from the entire strategy. Given the assumptions above on phase-II experiments, it'll turn out we can directly rule out one class of induced strategies from the sender's consideration. The other set of induced strategies will need careful assessment that we present next.

\vspace{-4pt}
\begin{claim} \label{clm0}
    When the inequalities of the assumption $q_{A1}\geq q_{A2}$ and $q_{B1}\geq q_{B2}$ in Lemma \ref{clm_asp} are strict, for any phase-II experiment $E_X \in \{E_A,E_B\}$), taking action $\phi_1$ when $E_X$ fails but taking action $\phi_2$ when $E_X$ passes is not an incentive compatible strategy for the receiver for any interim belief $~\mathbb{P}(\theta_1|E_X)\in [0,1]$.
\end{claim}

The above claim can be verified by comparing the posterior belief $\mathbb{P}(\theta_1|\omega)$ of each possible outcome $\omega\in \{\omega_{XP},\omega_{XF}\}$ and the receiver's corresponding best response. 
Therefore, upon the outcome of a phase-I experiment being revealed (to be either $\omega_A$ or $\omega_B$), the sender only has three different classes of ``\textit{induced strategies}'' by which to persuade the receiver in phase II:
\vspace{6pt}\\
($\alpha_X$) Suggest action $\phi_1$ only when the phase-II experiment outcome is a pass; 
\\
($\beta_X$) Suggest action $\phi_1$ no matter the result of phase-II experiment; and
\\
($\gamma_X$) Suggest action $\phi_2$ irrespective of the result of phase-II experiment, which is equivalent to not persuading the receiver to take the sender-preferred action. 
\vspace{6pt}

Given these three classes of induced strategies and the freedom to choose different induced strategies based on the phase-I experiment's outcome, the sender can use any combination of these $3^2$ choices to form a set of strategies $\mathcal{S}$. To simplify the representation, we use $(c_A,d_B)$, $c,d\in \{\alpha,\beta,\gamma \}$ to represent a ``\textit{type of strategy}'' of the sender. Note that to specify a strategy $S$ within the set of strategies, i.e., $S\in \mathcal{S}$, the probability pair $(p_1,p_2)$ has to be determined first. Before we analyze the different strategies, we discuss the relationship between the given phase-II experiments, induced strategies, and the incentive-compatibility requirements 
from the sender's side (to avoid profitable deviations by the receiver). 
To avoid ambiguity, hereafter, when we mention incentive compatibility/incentive compatible requirements/IC strategies, we mean the condition/requirements/strategies of a sender's commitment satisfying the following statement:
for every possible realized signal under this commitment, the receiver taking the sender-suggested action is incentive-compatible.

\vspace{-12pt}
\subsection{Constraints given by phase-II experiments} \label{sec:IC_in_phase2}
\vspace{-4pt}

By her choice of the experiment in phase I, the sender decides how to split the prior into the interim beliefs for the two experiments available in phase-II. The resulting interim-beliefs then lead to certain induced strategies at stage-II being applicable, i.e., incentive compatible (for the receiver). In other words, the probability pair $(p_1,p_2)$ must make each (applied) induced strategy yield the maximum utility for the receiver. These requirements constrain the sender's choice of $(p_1,p_2)$, and the sender needs to account for the (reduced) choice while deciding the split of the prior. Table~\ref{table:primaryIC} summarizes the impact in terms of the parameters of the phase-I experiment via primary
requirements on $(p_1,p_2)$ driven by the incentive compatibility while using each class of induced strategies. Hereafter, when we use IC requirements without additional specification, we mean primary IC requirements. From the entries in the table, it is clear that the phase-II experiments (indirectly) limit the sender's strategy selection where this limitation arises due the receiver's IC requirements for each induced strategy (when that induced strategy is used).  

\begin{table}[b] \vspace{-16pt}
\caption{IC requirements of the sender's commitment based on the induced strategy}
\vspace{-4pt}
\begin{tabular}{|c|l|c|l|}
\hline
 Induced strategy& Primary IC requirement &  Induced strategy& Primary IC requirement\\ \hline
$\alpha_A$ & $p_1\geq \frac{1-p}{p}\frac{q_{A2}}{q_{A1}}p_2$ &
$\alpha_B$ &  $1-p_1\geq \frac{1-p}{p}\frac{q_{B2}}{q_{B1}}(1-p_2)$ \\ \hline
$\beta_A$ & $p_1\geq \frac{1-p}{p}\frac{1-q_{A2}}{1-q_{A1}}p_2$ &
$\beta_B$ &  $1-p_1\geq \frac{1-p}{p}\frac{1-q_{B2}}{1-q_{B1}}(1-p_2)$ \\ \hline
$\gamma_A$ &  
$p_1\leq \frac{1-p}{p}\frac{q_{A2}}{q_{A1}}p_2$&
$\gamma_B$ & 
$1-p_1\leq \frac{1-p}{p}\frac{q_{B2}}{q_{B1}}(1-p_2)$
\\ \hline
\end{tabular}
\label{table:primaryIC} 
\vspace{-16pt}
\end{table}
With this in mind, the sender's experiment design in phase I, essentially, is to select between different combinations of these induced strategies such that each induced strategy satisfies the constraint listed in Table~\ref{table:primaryIC}. Hence, we next seek to understand how these IC constraints collectively determine the sender's strategy selection. To answer this, we first discuss the relationship between induced strategies, IC requirements, and the sender's expected utility.

From the sender's perspective, each induced strategy and its corresponding signals provide a path to persuade (or dissuade) the receiver to take action $\phi_1$. Since the sender's objective is to maximize the probability that action $\phi_1$ is taken, she would like to use the ``most efficient"\footnote{The efficiency of a strategy is defined as $\frac{\mathbb{P}(\phi_1|\text{interim belief, the induced strategy used})}{\mathbb{P}(\theta_1|\text{interim belief, the induced strategy used})}$.} pair of induced strategies to persuade the receiver\footnote{When the prior falls in the region where the optimal signaling strategy is non-trivial.}. 
To better understand the ``efficiency" of induced strategies, we evaluate each induced strategy under a given phase-II experiment $E_X$:\vspace{4pt}\\
$\bullet$\   \textbf{$\alpha_X$ strategy:} To persuade receiver to take action $\phi_1$ via this induced strategy, the sender needs to ensure that $\mathbb{P}(\theta_1|\omega_{XP})\geq \frac{1}{2}$. Hence, the interim belief $\mathbb{P}(\theta_1|E_X)$ must satisfy $\mathbb{P}(\theta_1|E_X) \in \left[\frac{q_{X2}}{q_{X1}+q_{X2}}, \frac{1-q_{X2}}{2-q_{X1}-q_{X2}}\right]$, otherwise a commitment using $\alpha_X$ induced strategy will never be incentive-compatible. From the sender's perspective, the most efficient strategy to persuade the receiver using $\alpha_X$ induced strategy is to design the phase-I experiment such that $\mathbb{P}(\theta_1|E_X)=\frac{q_{X2}}{q_{X1}+q_{X2}}$. At this interim belief, the sender experiences a relative expected utility 
$2q_{X1}$ (with respect to the prior). When $\mathbb{P}(\theta_1|E_X) \in \left(\tfrac{q_{X2}}{q_{X1}+q_{X2}}, \frac{1-q_{X2}}{2-q_{X1}-q_{X2}}\right)$, the sender's marginal expected utility when the interim belief increases is $q_{X1}-q_{X2}$.\vspace{4pt}\\
$\bullet$\    \textbf{$\beta_X$ strategy:} To persuade receiver to take action $\phi_1$ with this induced strategy the sender needs to ensure that both inequalities $\mathbb{P}(\theta_1|\omega_{XP})\geq \frac{1}{2}$ and $\mathbb{P}(\theta_1|\omega_{XF})\geq \frac{1}{2}$ hold. Given the assumption in Lemma \ref{clm_asp}, namely $q_{X1}\geq q_{X2}$, the only constraint that can be tight is $\mathbb{P}(\theta_1|\omega_{XF})\geq \frac{1}{2}$. Hence, IC commitments using $\beta_X$ induced strategy exist only when the interim belief $\mathbb{P}(\theta_1|E_X) \geq \frac{1-q_{X2}}{2-q_{X1}-q_{X2}}$. From the sender's perspective, the most efficient strategy to persuade the receiver using a $\beta_X$ induced strategy is to design the phase-I experiment such that $\mathbb{P}(\theta_1|E_X)=\frac{1-q_{X2}}{2-q_{X1}-q_{X2}}$ with the resulting relative expected utility $1+\frac{1-q_{X1}}{1-q_{X2}}$. Unlike an $\alpha_X$ induced strategy where the sender still gets a positive utility gain when the interim belief increases, for a $\beta_X$ induced strategy the sender's marginal expected utility gain when the interim belief increases is $0$ when $\mathbb{P}(\theta_1|E_X)>\frac{1-q_{X2}}{2-q_{X1}-q_{X2}}$. 
\vspace{4pt} \\
$\bullet$\    \textbf{$\gamma_X$ strategy:} Since the sender suggests the receiver to take action $\phi_2$ in this strategy, the sender's expected utility is $0$ when using this induced strategy.

According to the discussion above, it is clear that the sender will not use the set of strategies corresponding to $(\gamma_A, \gamma_B)$ unless the prior $p=0$.  
Besides, we know that different induced strategies provide different relative expected utility to the sender. When induced strategies are used in the most efficient manner, the relative expected utility under a $\alpha_X$ induced strategy is at most $2q_{X1}$, and the average expected utility under a $\beta_X$ induced strategy is at most $1+\tfrac{1-q_{X1}}{1-q_{X2}}$.

Since these two values capture the best scenario that the sender can achieve by tailoring the interim belief under the given experiment, we define this pair of ratios, $(2q_{X1},1+\tfrac{1-q_{X1}}{1-q_{X2}})$ as a function of (the given) experiment, denoted by $PerP(E_X)$; this pair is called the \textit{persuasion potential}. 

\vspace{-2pt}
\begin{definition}
     Given an experiment $E_X=(q_{X1},q_{X2})$, the persuasion potential of this experiment, $PerP(E_X)$, is the pair $\Big(2q_{X1}, 1+\frac{1-q_{X1}}{1-q_{X2}}\Big)$.
\end{definition}
\vspace{-4pt}
To provide some insights on the importance and use of the persuasion potential we preview Corollary \ref{cor:perppowerful}. Corollary \ref{cor:perppowerful} states that the sender only uses induced strategies in the most efficient manner, i.e., $\mathbb{P}(\theta_1|E_X)=\tfrac{q_{X2}}{q_{X1}+q_{X2}}$ when an induced strategy $\alpha_X$ is used and $\mathbb{P}(\theta_1|E_X)=\tfrac{1-q_{X2}}{2-q_{X1}-q_{X2}}$ when an induced strategy $\beta_X$ is used. Thus, the persuasion potential can simplify the sender's search for the optimal signaling strategy. When a particular induced strategy is used in the most efficient manner described in the above parameter, the interim belief is now determined. Therefore, the sender does not need to search for the optimal signaling strategy from the whole set of IC strategies but only needs to search from a small number of strategies that generate the particular interim beliefs.

\vspace{-12pt}
\subsection{Persuasion ratio and the optimal signaling structure} \label{sec:optsig}
\vspace{-6pt}
Since the sender wants to maximize the total probability of action $\phi_1$, she needs to compare different sets of strategies formed by different pairs of induced strategies. To compare 
each set of strategies, we introduce the persuasion ratio of a set of strategies for a given value of the prior.
\vspace{-4pt}
\begin{definition} \label{prcurve}
Given a set of incentive-compatible strategies $\mathcal{S}$,  e.g., $\mathcal{S}=(c_A,d_B)$ with $c,d\in \{\alpha,\beta,\gamma \}$ which satisfying IC requirements, the persuasion ratio of the set of strategies $\mathcal{S}$ is the maximum total probability of action $\phi_1$ is taken (under a strategy within the set) divided by the prior $p$, $PR(\mathcal{S},p)=\max_{S\in \mathcal{S}} \frac{\mathbb{P}(\phi_1|S)}{p}$.
\end{definition}
\vspace{-4pt}

Careful readers may notice that if we multiply the persuasion ratio with the prior, the value will be the (maximum) expected utility the sender can achieve from the given set of strategies. Since the sender's expected utility is monotone increasing in the prior $p$ regardless of which set of strategies the sender adopts, the persuasion ratio under a given prior can be viewed as the relative utility gain this set of strategies can offer to the sender. Hence, given a specific prior, if a set of strategies has a higher persuasion ratio with respect to another set of strategies, the sender should use a strategy in the former instead of the latter.

According to this discussion, we can draw a persuasion ratio curve for each set of strategies as the prior is varied in $[0,1]$. Abusing notation, we represent the persuasion ratio curve by $PR(\mathcal{S})$. It may appear that an optimization needs to be carried out for each value of the prior. However, structural insights presented in the following two lemmas considerably simplify the analysis. Properties presented in Lemma \ref{lem:tightness} narrow down the space where the sender needs to search for the optimal signaling strategy. This allows us to depict persuasion ratio curves $PR(S)$ for some basic strategies. On top of that, Lemma \ref{lem:conveshull} provides a systematic approach to derive persuasion ratio curves for all types of strategies.

\vspace{-6pt}
\begin{lemma} \label{lem:tightness}
    Given a type of strategy $\mathcal{S}$ and a prior $p$, there exists a (sender's) optimal strategy $S \in \mathcal{S}$ which satisfies one of the following two conditions:
    \vspace{-4pt}
    \begin{enumerate}
        \item  At least one IC requirement of the constituent induced strategies is tight;
        \item There is a signal that will be sent with probability $1$ under $S$.
    \end{enumerate}
\end{lemma}
\vspace{-6pt}
Before discussing the simplifications that Lemma \ref{lem:tightness} yields in terms of the key properties for solving the problem, we give an intuitive outline of the proof of Lemma \ref{lem:tightness}. When the IC requirements of the two induced strategies are not tight and both signals are sent with non-zero probability in a strategy $S$, the sender can increase her expected utility by slightly raising the probability of the signal with a higher persuasion ratio (and adjust the probability of the other signal to respect the prior) to form a strategy $S_+$. The sender can keep doing this `slight' modification of her strategies until either one of the IC conditions is satisfied or the signal is sent with probability one.

Given this lemma, the persuasion ratio curve of the following types of strategies: $(\alpha_A,\gamma_B)$, $(\beta_A,\gamma_B)$, $(\gamma_A,\alpha_B)$, $(\gamma_A,\beta_B)$ can be determined immediately since the IC requirement can never be tight for the $\gamma$ class induced strategy. For the remaining four types of strategies: $(c_A,d_B)$, $c,d \in \{\alpha,\beta\}$, the following lemma aids in solving for the strategy meeting the persuasion ratio without a point-wise calculation. As a preview of result of Lemma \ref{lem:optbny}, once we derive the persuasion ratio curve for each type of strategies via Lemma \ref{lem:conveshull}, we can immediate identify the optimal signaling strategy by overlaying those curves in one figure.

\vspace{-4pt}
\begin{lemma} \label{lem:conveshull}
    Given the persuasion ratio curves of the types of strategies $(c_A,\gamma_B)$ and $(\gamma_A,d_B)$, denoted by $PR((c_A,\gamma_B),p)$ and $PR((\gamma_A,d_B),p)$, respectively, the persuasion ratio curve of the set of incentive compatible strategies $(c_A,d_B)$, denoted by $PR((c_A,d_B),p)$, is the generalized concave hull~\footnote{If $PR((c_A,\gamma_B),\cdot)$ and $PR((\gamma_A,d_B),\cdot)$ were the same function $f(\cdot)$, then this would be its convex hull.} of the functions
    $PR((c_A,\gamma_B),p)$ and $PR((\gamma_A,d_B),p)$ when $\mathcal{S}_{(c_A,d_B)}(p) \neq \emptyset$: 
    \begin{equation}\label{eq:prcomb}
        PR((c_A,d_B),p)=\max_{\substack{x,u,v\in [0,1], \\ xu+(1-x)v=p }}xPR((c_A,\gamma_B),u)+(1-x)
        PR((\gamma_A,d_B),v)), \vspace{-8pt}
    \end{equation}
     where $\mathcal{S}_{(c_A,d_B)}(p)$ is the set of IC strategies of $(c_A,d_B)$ at $p$. \vspace{-4pt}
\end{lemma}

The proof of Lemma \ref{lem:conveshull} uses the structure of the sender's expected utility when no $\gamma$ induced strategy is used in the types of strategies employed. The sender's expected utility function under $(c_A,d_B)$ at prior $p$ can be represented as a linear combination of her utility function under $(c_A,\gamma_B)$ at prior $u$ and her utility function under $(\gamma_A,d_B)$ at prior $v$, then the optimization problem of solving the optimal phase-I experiment parameters $(p_1,p_2)$ can be transformed to the maximization problem in the statement of Lemma \ref{lem:conveshull}.
As we have discussed the means to determine each type of strategy's persuasion ratio curve, Lemma \ref{lem:optbny} illustrates the persuasion ratio curve of the optimal signaling strategy.

\vspace{-4pt}
\begin{lemma} \label{lem:optbny}
The persuasion ratio curve of the optimal signaling strategy $PR^*(p)$ is the upper envelope of the different types of strategies' persuasion curves. Further, the optimal signaling strategy (under a given prior) is the strategy that reaches the frontier of the persuasion ratio curve (at that prior). \vspace{-4pt}
\end{lemma}

    Since a higher persuasion ratio indicates a higher (sender's) expected utility for every given prior, the sender will choose the upper envelope of the persuasion curves of the different types of strategies. Because the set $\mathcal{S}_{(c_A,d_B)}(p)$ could be empty for some prior values with a corresponding persuasion ratio $PR(c_A,d_B),p)=0$, the main effort in proving Lemma \ref{lem:optbny} is to show the existence of an incentive-compatible 
commitment on the frontier of the persuasion ratio curve at every possible prior.
Finally, once the persuasion ratio curve of the optimal signaling strategy is determined, we can immediately infer an optimal signaling strategy $S^*$ under a specific prior.

For a two-phase trial or to solve the last two phases of a trial with more than two phases studied in Section \ref{sec:multiphase}, the following corollary can further simplify the sender's optimization procedure.
\vspace{-4pt}
\begin{corollary} \label{cor:perppowerful}
Let\/ $\Pi^*(p)$ represents the optimal signaling strategy at prior $p$. If\/ $\mathbb{P}(\phi_1|\Pi^*(p))<1$, then the following two statements are true: \vspace{-4pt}
\begin{enumerate}
    \item When $\alpha_X$ is used in $\Pi^*(p)$, the interim belief is $\mathbb{P}_{\Pi^*(p)}(\theta_1|E_X)=\tfrac{q_{X2}}{q_{x1}+q_{X2}}$.
    \item When $\beta_X$ is used in the optimal signaling strategy\/ $\Pi^*(p)$, the interim belief is\/  $\mathbb{P}_{\Pi^*(p)}(\theta_1|E_X)=\frac{1-q_{X2}}{2-q_{x1}-q_{X2}}$.
\end{enumerate}
\end{corollary}
\vspace{-4pt}

With Corollary \ref{cor:perppowerful}, the Equation~\eqref{eq:prcomb} in Lemma \ref{lem:conveshull} reduces to a linear equation. Hence, the comparison in Lemma \ref{lem:optbny} and the computation in Lemma \ref{lem:conveshull} can be reduced to a comparison of the (unique) corresponding IC strategies (if one exists) under the interim belief listed in Corollary\ref{cor:perppowerful} for different types of strategies.
\vspace{-12pt}
\subsection{Comparison with classical Bayesian persuasion strategies}
\vspace{-4pt}
Given the optimal signaling strategy derived in Lemma \ref{lem:optbny}, one natural follow-up question is the quantification of the sender's utility improvement obtained by adopting the optimal signaling strategy in comparison to using strategies structurally similar to the optimal strategies in classical Bayesian persuasion for a binary state of the world. 
Owing the page limit, we directly define a class of strategies structurally similar to the classical Bayesian persuasion strategy below and provide the justification in our online version \cite{our_arxiv}.

\vspace{-4pt}
\begin{definition}
With binary states of the world, a (binary-state) Bayesian persuasion (BBP) strategy is a strategy that ``mixes two possible states in one signal and reveals the true state on the other signal''. \vspace{-4pt}
\end{definition}
 


\begin{figure}[h] \vspace{-18pt}
    \centering
    \includegraphics[width=0.8\textwidth]{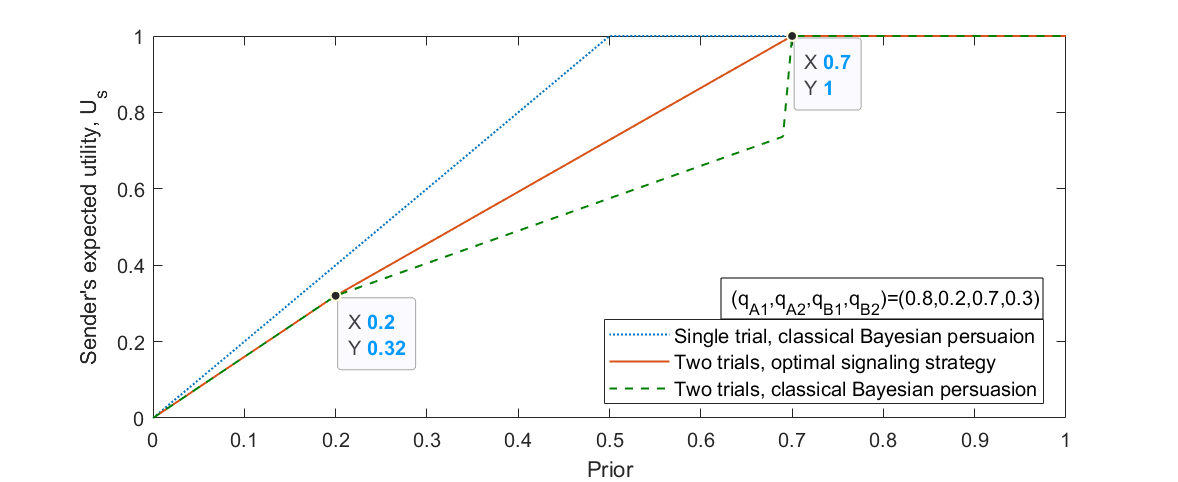} \vspace{-12pt}
    \caption{Sender's utility under different problem settings and strategies}
    \label{fig:BP2trials} \vspace{-12pt}
\end{figure}

Given the model defined in Section \ref{sec:basicmodel}, a BBP strategy is forced to use at least one $\gamma_X$ induced strategy\footnote{Because using either $\alpha_X$, $\beta_X$ requires a mixture of two possible states in one signal.}. Given a fixed type of strategy, e.g., $(\alpha_A,\gamma_B)$, an optimal BBP strategy using this type of strategy can be solved by the concavification approach after the calculation of the sender's expected utility under interim beliefs. After solving the optimal BBP strategy of a given strategy type via concavification respectively, the optimal BBP strategy is the strategy in the set of $\{(\alpha_A,\gamma_B),(\beta_A,\gamma_B),(\gamma_A,\alpha_B),(\gamma_A,\beta_B)\}$ which yields the highest expected utility for the sender. Figure \ref{fig:BP2trials} plots the sender's expected utility for the optimal signaling strategy and the optimal BBP strategy under a given pair of phase-II experiments: $(q_{A1},q_{A2})=(0.8,0.2), (q_{B1},q_{B2})=(0.7,0.3)$. The blue line in Figure 2 is the benchmark of the sender's maximum expected utility in a single-phase scenario where the sender chooses the experiment.  Note this would also be the optimal performance if one of phase-II trials were changed to a trivial experiment. As we can see, the sender's expected utility is lowered owing to the determined phase-II experiments. For low-priors, the optimal signaling strategy derived in Section \ref{sec:optsig} and the optimal BBP strategy give the sender the same expected utility\footnote{Because the optimal signaling strategy in the low-prior region is $(\alpha_A,\gamma_B)$ here.}. However, as the prior increases, a utility gap between the optimal signaling strategy derived in Section \ref{sec:optsig} and the optimal BBP strategy appears and then increases until the receiver will take $\phi_1$ with probability one. The utility gap starts when the optimal signaling strategy uses strategies $(\alpha_A,\beta_B)$ or $(\alpha_B,\beta_A)$ which are not considered in BBP strategies.

\vspace{-18pt}
\section{Binary-outcome Experiments in Multi-phase trials}\label{sec:multiphase}
\vspace{-8pt}

This section generalizes the structural results in Section \ref{sec:BOE} to multi-phase trials. First, we generalize the model in Section \ref{sec:basicmodel} to multi-phase trials and then propose a dynamic programming algorithm to solve for the optimal signaling strategy. The state for the dynamic program will be the interim belief on the state of the world that results at any node in the extensive-form delineation of the problem. As the belief at each level is determined based on the actions in earlier stages (if any), in the backward iteration procedure, we will determine the optimal choice of experiments by the sender (if there is a choice) for any possible interim belief. In this dynamic programming, there is only a terminal reward that arises from the receiver's action based on the outcome of the final trial and based on the receiver's resulting posterior beliefs.
\vspace{-8pt}
\subsection{Model of binary-outcome experiments in multi-phase trials}
\vspace{-4pt}
There are $N$ phases in a trial where one binary-outcome experiment will be conducted in each phase. However, as in the two-phase-trial settings, the specific experiment conducted in each phase is determined by the earlier phases' outcomes. Therefore, we can model an $N$-phase trial by a height-$N$ binary tree where each leaf node represents an outcome revealed to the receiver, and each non-leaf node represents an experiment. With this binary tree, experiment $E_{i,j}$ represents the $j^{th}$ experiment to be conducted at level $i$. When $j$ is odd, an experiment $E_{i,j}$ will be conducted only if the experiment $E_{i-1,(j+1)/2}$ is conducted and passed. Similarly, when $j$ is even, an experiment $E_{i,j}$ will be conducted only if the experiment $E_{i-1,j/2}$ is conducted but it fails. In this binary tree, some experiments, e.g, $E_{i,j}$, are determined. However, some experiments, e.g., $E_{k,l}$, can be designed by the sender; all the parameters are chosen before any experiment is conducted and are common knowledge. In such experiments, the sender can choose a probability pair, e.g., $(p_{kl1},p_{kl2})\in[0,1]^2$. In contrast to the model defined in Section \ref{sec:basicmodel}, here determined experiments and sender-designed experiments can be at any level of the tree with the placement arbitrary but carried out before the sender receives her signals. In other words, unlike the model in Section \ref{sec:basicmodel}, a sender may be able to design an experiment at final level (phase $N$) owing to a determined experiment outcome at phase $N-1$. 

In this model, the prior, the experiments that are determined (their location on the tree and their parameters), and the sender-designed experiments (their location on the tree) are common knowledge. The sender has to design all her experiments simultaneously and before the game starts (when the state is realized); the designed experiments' parameters are then revealed to the receiver (again before the game starts). Given the experiments designed by the sender and the realized outcome of a sequence of experiments, the receiver will take an action to guess the true state of the world. For simplicity of analysis, we keep the sender and the receiver's utilities the same as in Section \ref{problem}. 
Then, the sender's objective is to jointly design the set of experiments that she has the flexibility to choose to maximize her expected utility, which is nothing but the probability of the receiver taking action $\phi_1$. Before proceeding, we point out that this model can be easily generalized to unbalanced binary trees straightforwardly by adding dummy nodes with determined trivial experiments defined in Section \ref{sec:sde} to construct an equivalent balanced binary tree.
\vspace{-12pt}
\subsection{Determined versus sender-designed experiments}
 \vspace{-4pt}
Given the model, the sender can manipulate the phase-$K$ interim belief only when designing an experiment at phase $K-1$. If an experiment at phase $K-1$ is determined, then the phase-$K$ interim belief is a function of interim belief at phase $K-1$. Therefore, figuring out how these two types of experiments, determined and sender-designed experiments, will influence every given phase's interim belief is the key to solving for the optimal signaling strategy. We start by noting that if the posterior belief at a leaf node is given, then the receiver's action is determined - he will take the action with the highest posterior probability unless there is a tie, in which case he is indifferent and will follow the sender's recommendation. Therefore, we can use backward iteration and the principle of optimality to determine the optimal signaling. We start by considering the last 
phase's experiments when the sender can design them. 

 \vspace{-15pt}
\subsubsection{Experiments at phase $N$}
 Recall the result we have discussed in Section \ref{sec:twophase}, a determined experiment in the last phase (phase-II in Section \ref{sec:twophase}) limits the sender's strategy choice to one of three induced strategies. Besides, the best scenario that the sender can achieve via using these induced strategies (without violating the IC requirement) is captured by the persuasion potential of the determined experiment. However, when there is a sender-designed experiment at phase $N$ and the interim belief\footnote{See the corresponding footnote in \cite{our_arxiv} for the discussion of the interim belief $\tilde{p}>\frac{1}{2}$.} $\tilde{p}\leq \frac{1}{2}$, the sender can always design an experiment which makes two states equally likely when this experiment passes and reveal the less-preferred state (by sender) when it fails. If we cast this sender-designed experiment in terms of a determined experiment, the sender-designed experiment will have a persuasion potential $(2,2)$
 \footnote{See the corresponding footnote in \cite{our_arxiv} for the detailed derivation.}. Thus, no matter the type of experiment at phase $N$, we can capture the sender's optimal set of induced strategies via a persuasion potential.
\subsubsection{Experiments in phase $N-1$} In the second-last phase, results in Section \ref{sec:optsig} describe a sender-designed experiment's role in the optimal signaling strategy: \textit{pick the strategy on the frontier of all persuasion-ratio curves}. However, if the experiment is determined in the second-last phase, an additional constraint on the interim belief between the second-last phase and the last phase is enforced. That is to say, the set of (feasible) strategies will shrink. 
Fortunately, after enforcing the constraints, the process of searching for the optimal signaling strategy under a determined experiment is the same as the sender-designed experiment, i.e., pick the strategy in the frontier of all persuasion ratio curves. Therefore, at each possible branch of phase $N-1$, we can plot an optimal persuasion ratio curve capturing the sender's optimal signaling strategy at phase $N-1$ and phase $N$.
  \setlength{\textfloatsep}{0pt}
    \begin{algorithm}[t]
\caption{Dynamic programming approach for multi-phase trials}  \label{alg:DP}
\SetKwInput{KwInput}{Input}                
\SetKwInput{KwOutput}{Output}              
\DontPrintSemicolon
  \KwInput{The set of determined experiments $\mathbf{E_D}$, the binary tree structure}
  \KwOutput{The optimal persuasion ratio curve}
  \textbf{1.} For each experiment at phase $N$, $E_{N,i}~i\in\{1,...,2^N\}$, solve its persuasion potential $Prep(E_{N,i})$ \\
  \textbf{2.} For each experiment at phase $N-1$, $E_{N-1,i}~i\in\{1,...,2^{N-1}\}$, find the optimal persuasion ratio curve using $(Prep(E_{N,2i-1}),Prep(E_{N,2i}))$. \\
  \textbf{3.} K=N-2 \\
  \textbf{4.} \While{$K>0$} 
  {For each experiment at phase $K$, $E_{K,i}~i\in\{1,...,2^{K}\}$, find the optimal persuasion ratio curve using equation (\ref{eqn:2}) or Claim \ref{clm2}\\
  K=K-1}
  \textbf{5.} Return the optimal persuasion ratio curve at phase $1$
\end{algorithm}
\vspace{-8pt}
\subsubsection{Experiments in earlier phases}
\vspace{-4pt}
Now we consider experiments in earlier phases. When we have a determined experiment in phase-$K$, e.g., $E_{k,i}=(q_{Ki1},q_{Ki2})$, and we have solved the optimal persuasion ratio curves of its succeeding phase ($(K+1)$), i.e., $PR^*_{K+1,2i-1}(p)$ and $PR^*_{K+1,2i}(p)$, then the optimal persuasion ratio curve at this determined phase-$K$ experiment $E_{k,i}$ is just a linear combination of $PR^*_{K+1,2i-1}(p)$ and $PR^*_{K+1,2i}(p)$ can be written as follows: \vspace{-6pt}
\begin{eqnarray}
    PR^*_{K,i}(p)&=&(pq_{Ki1}+(1-p)q_{Ki2})PR^*_{K+1,2i-1}\Big(\tfrac{pq_{Ki1}}{pq_{Ki1}+(1-p)q_{Ki2}}\Big)+(p(1-q_{Ki1})\nonumber\\
    &+&(1-p)(1-q_{Ki2}))PR^*_{K+1,2i}\Big(\tfrac{p(1-q_{Ki1})}{p(1-q_{Ki1})+(1-p)(1-q_{Ki2})}\Big) \label{eqn:2} \vspace{-6pt}
\end{eqnarray}

For a sender-designed experiment at phase K, e.g., $E_{K,j}$, if we have already solved the optimal persuasion ratio curves of its succeeding phase-$(K+1)$ experiments $E_{K+1,2j-1}$ and $E_{K+1,2j}$, the sender's best design at $E_{K,j}$ is to find a linear combination of $PR^*_{K+1,2j-1}(p)$ and $PR^*_{K+1,2j}(p)$ which yield the highest persuasion ratio for every phase-$K$ interim belief $p$. Since the persuasion ratio curve is monotone decreasing in the belief, the optimal persuasion ratio curve can be constructed similar to Lemma \ref{lem:conveshull} as shown in Corollary \ref{clm2}.
\vspace{-4pt}
\begin{corollary} \label{clm2}
    Given two persuasion ratio curves at phase $K+1$, $PR^*_{K+1,2j-1}(p)$ and $PR^*_{K+1,2j}(p)$, the optimized persuasion ratio curve $PR^*_{K,j}(p)$ at phase $K$ is the maximum convex combination of $PR^*_{K+1,2j-1}(p)$ and $PR^*_{K+1,2j}(p)$, i.e.,
    \begin{eqnarray} \label{eq:prcomb_nonbny}
        PR^*_{K,j}(p)=\max_{x,u,v\in[0,1],xu+(1-x)v=p}xPR^*_{K+1,2j-1}(u)+(1-x)
        PR^*_{K+1,2j}(v)
    \end{eqnarray} 
\end{corollary}
\vspace{-12pt}
\subsubsection{Non-binary Outcome Experiments} \label{sec:nonbinaryexp}
When the experiments have non-binary outcomes, the same approach derived above works with an increased number of phases (if complexity is not an issue). 

For general non-binary experiments, see the proof of Lemma \ref{clm:nbnry} in \cite{our_arxiv} for a detailed construction from non-binary to binary experiments.
\vspace{-4pt}
\begin{lemma}\label{clm:nbnry}
Given a non-binary experiment $E=\{q_{1,1},...,q_{1,n};q_{2,1},...,q_{2,n}\}$, we can replace it by $\lceil \log_2n \rceil$ levels of binary outcome experiments.
\end{lemma}
\vspace{-12pt}
\subsection{Multi-phase model and classical Bayesian persuasion}
\vspace{-4pt}
At the end of this section, we mention a class of special multi-phase trials where the sender's expected utility under the optimal signaling strategy is equivalent to utility obtained from a single-phase Bayesian persuasion model. Inspired by the two-phase example with a trivial experiment in \ref{lem:screening}, the sender can implement a signaling strategy similar to single-phase Bayesian persuasion when there exists a trivial experiment in the last phase and she can design experiments in earlier phases. When the sender can design all earlier phases, she can voluntarily reduce the signal space in effect via designing the experiment $E_{i,j}$ to be $E_{i,j}=(1,1)$ or $E_{i,j}=(0,0)$, i.e., a non-informative experiment. By doing this, the sender can reduce the multi-phase trial to an equivalent two-phase trial model and then a straightforward extension of Lemma \ref{lem:screening} will hold when there exists a trivial experiment in the last phase. 
The following Lemma~\ref{lem:nphasebp} further generalizes the class of multi-phase models where the sender has the same expected utility as a single-phase Bayesian persuasion problem with a necessary pruning process defined in Definition \ref{dfn:prune}. Owing to the page limit, explanations and some preliminary analysis about the robustness of signaling strategies under small perturbations from trials satisfying Lemma \ref{lem:nphasebp} is only available in our online version \cite{our_arxiv}.

\vspace{-4pt}
\begin{definition} \label{dfn:prune}
     Given an N-phase trial model $M$, a pruned N-phase trial model $Prun(M)$ is a model which recursively replaces every subtree of $M$ by a revealing experiment $E_{\theta}=(1,0)$ if the subtree satisfies the following condition, starting from the leaves: the (sub)root of this subtree has a trivial (determined) experiment $E_X$ with at least one of its succeeding experiments non-trivial (but determined).
\end{definition} 
Note that the pruned tree will potentially be unbalanced.

\begin{lemma} \label{lem:nphasebp}
Given an $N$-phase trial $M$ with binary-outcome experiments, if there exists a pruned N-phase trial model $Prun(M)$ such that the following two conditions hold, then the sender's expected utility is given by an equivalent single-phase Bayesian persuasion model.
\begin{enumerate}
        \item For every non-trivial determined experiment, its sibling is either a trivial or a sender-designed experiment.
        \item There exists a least one sender-designed experiment in each (from root to leaf) experiment sequence of $Prun(M)$ .
\end{enumerate}
\end{lemma}

\vspace{-18pt}
\bibliographystyle{splncs04}
\bibliography{X_Ref_incentivizinginfo}
\end{document}